\documentclass[trackchanges, twocolumn]{aastex701}

\usepackage{commath}
\usepackage{txfonts}
\usepackage{subfig}
\usepackage{upgreek}
\usepackage{threeparttable}
\usepackage{booktabs} 

\newcommand{\ze}{PKS\,0735+178}
\newcommand{\beam}{0.25\,\textrm{mas}}
\newcommand{\rms}{1\,\textrm{mJy/beam}}

\begin{document}

\title{Template \aastex v7.0.1 Article with Examples\footnote{Footnotes can be added to titles}}

\title{Polarisation as a probe of neutrino emission from blazars}

\correspondingauthor{G.~F. Paraschos}
\author[orcid=0000-0001-6757-3098,gname='Georgios Filippos', sname='Paraschos']{G.~F. Paraschos}
\affiliation{Max-Planck-Institut f\"ur Radioastronomie, Auf dem H\"ugel 69, D-53121 Bonn, Germany}
\email[show]{gfparaschos@mpifr-bonn.mpg.de}

\author[orcid=0000-0002-1209-6500,gname='Efthalia', sname='Traianou']{E. Traianou}
\affiliation{Interdisziplinäres Zentrum für wissenschaftliches Rechnen (IWR), Ruprecht-Karls-Universität Heidelberg, Im Neuenheimer Feld 205, 69120, Heidelberg, Germany}
\affiliation{Max-Planck-Institut f\"ur Radioastronomie, Auf dem H\"ugel 69, D-53121 Bonn, Germany}
\email[]{traianouthalia@gmail.com}

\author[orcid=0009-0003-8342-4561,gname='Lena Carolin', sname='Debbrecht']{L.~C. Debbrecht}
\affiliation{Max-Planck-Institut f\"ur Radioastronomie, Auf dem H\"ugel 69, D-53121 Bonn, Germany}
\email[]{ldebbrecht@mpifr-bonn.mpg.de}

\author[orcid=0000-0001-9200-4006,gname='Ioannis', sname='Liodakis']{I. Liodakis}
\affiliation{Institute of Astrophysics, Foundation for Research and Technology-Hellas, GR-71110 Heraklion, Greece}
\affiliation{Max-Planck-Institut f\"ur Radioastronomie, Auf dem H\"ugel 69, D-53121 Bonn, Germany}
\email[]{yannis.liodakis@gmail.com}

\author[orcid=0000-0001-9503-4892,gname='Eduardo', sname='Ros']{E. Ros}
\affiliation{Max-Planck-Institut f\"ur Radioastronomie, Auf dem H\"ugel 69, D-53121 Bonn, Germany}
\email[]{eros@mpifr-bonn.mpg.de}

\begin{abstract}

The source of extragalactic neutrinos in the TeV--PeV range is a matter of very active research, with blazar jets having been postulated to be the origin of at least some of the detections.
The blazar PKS\,0735+178 is a prominent example; during its multi-band flare in late 2021 a neutrino event was reported by four observatories, with its origin consistent with the direction of that source.
While no new jet component was observed to be ejected during that narrow time-frame, our analysis shows that a propagating shock front originating from the core region was the likely source of the multi-band flare, using very-long-baseline interferometry images of PKS\,0735+178 in polarised light.
Taken together, our findings are suggestive of a coherent scenario in which the shock may contribute to the acceleration of protons, with the target photons potentially originating either from the ambient medium surrounding the jet or from proton synchrotron radiation.
The necessary conditions for neutrino emission via proton--photon interactions are, hence, present in this jet.
\end{abstract}


\section{Introduction}
The advent of extragalactic, neutrino-related multi-messenger astronomy was ushered in with the announcement of the detection of PeV-energy neutrinos by the IceCube collaboration \citep{Aartsen13, Icecube13}.
Since then, numerous similar neutrino events have been reported \citep{Aartsen20}, spurning the community to investigate their potential origins.
As high-energy astrophysical sources are thought to produce them, various ones such as X-ray binaries \citep{Koljonen23}, active galactic nuclei (AGN), supernovae remnants, and tidal disruption events have been suggested \citep[see e.g.][for a review]{Boettcher22}.
AGN, and specifically blazars (an AGN sub-category with the jet pointed close to our line of sight) have turned out to be promising candidates for at least some of the neutrino events.
The smoking gun for the neutrino emission was originally considered to be a prominent flare in $\gamma$-rays, although later studies suggested that it might not be a necessary prerequisite for neutrino production \citep[e.g.][]{Franckowiak20, Garrappa24}
Similarly, studies by \citet{Plavin20}, \citet{Hovatta21}, and \citet{Kouch24} have attempted to statistically connect radio and optical flaring states to neutrino production but have not been conclusive. 
On the contrary, $\gamma$-ray flares have been statistically connected to rotations of the polarisation angle seen in optical \citep{Blinov15, Blinov18}.
Therefore, polarisation angle variations could be an additional probe in the blazar-neutrino connection.

Individual blazars have also been linked neutrino detections before.
Most famously, the blazar TXS\,0506+056 had a number of neutrino events related to its position on the sky \citep{Icecube18a, Icecube18b, Ansoldi18, Petropoulou20}, leading to numerous follow-up works which modelled its physical characteristics \citep[see][for a review]{Halzen22}.
Some other notable examples include PKS\,B1424$-$418 \citep{Kadler16}, PKS\,1502+106 \citep{Rodrigues21},  3HSP\,J095507.9+35510 \citep{Giommi20a}, and PKS\,1424+240 \citep{Padovani22, Kovalev25}.

Another source coincident with a neutrino event is \ze\ ($z=0.45$; \citealt{Nilsson12} and $\textrm{M}_\bullet = 1.89\times10^8 \textrm{M}_\odot$; \citealt{Gupta12}).
Intriguingly, the localisation area of the incoming signal includes \ze\ when taking into account systematics \citep{Giommi20b, Plavin20, Hovatta21}.
In the immediate time span around the neutrino arrival picked up by IceCube \citep[][December 8, 2021]{Icecube21}, three more neutrino detection instruments reported possible neutrino detections, namely Baikal-GVD \citep{Dzhilkibaev21}, Baksan Underground Scintillation Telescope \citep{Petkov21}, and KM3NeT \citep{Filippini22}).
At the same time, this blazar, which is a bright source in the northern radio sky, exhibited a flare in $\gamma$-ray, X-ray, ultraviolet, infrared, and optical bands \citep[e.g.][]{Acharyya23}.
Its characterisation as an intermediate-high-energy-peaked BL Lacertae object (IHBL) matches the other aforementioned neutrino emitter candidates TXS\,0506+056, PKS\,1424+240, as well as GB6\,J1542+6129 \citep{Padovani22}.
Furthermore, as pointed out by \cite{Britzen10}, who analysed centimetre VLBI epochs of the source between 1995 and 2008, \ze\ goes through phases of stationarity and twisted, superluminal motion.
 
In this work we showcase tentative evidence for an association between the neutrino emission and the magnetic field configuration of the jet in \ze\ as manifested in polarised light.
The paper is structured as follows: in Sect.~\ref{sec:Methods} we present the observations we used and the analysis we implemented; in Sect.~\ref{sec:Results} we showcase the results we obtained.
In Sect.~\ref{sec:Discussion} we discuss the implications of such a blazar-neutrino connection.
Finally, in Sect.~\ref{sec:Conclusions} we summarise our conclusions.
For all the calculations we adopt a $\Lambda$CDM cosmology with: $\Omega_\mathrm{M}=0.27$, $\Omega_{\Lambda}=0.73$, $H_{0}=71$\,km\,s$^{-1}$\,Mpc$^{-1}$ \citep{2009ApJS..180..330K}, which result in a luminosity distance of $D_{L}=2.53$\,Gpc. 
This implies a conversion factor of $\psi=5.57$\,pc/mas for the redshift $z=0.45$.

\section{Observations and data analysis}\label{sec:Methods}

To study the jet of \ze, we used a combination of its $\gamma$-ray light curve, optical electric vector position angles (EVPAs), its VLBI data, and the coincident neutrino detection.
Specifically, the track-like neutrino event IceCube-211208A (30\% probability of being astrophysical), which was detected by IceCube on the 8th of December 2021 (MJD 59554), at 172\,TeV \citep{Icecube21}.
Contemporary signals were also picked up by Baikal-GVD \citep[43\,TeV cascade event at 50\% probability of being astrophysical][]{Dzhilkibaev21},
Baksan Underground Scintillation Telescope \citep[muon neutrino
detection at $>1$\,GeV][]{Petkov21}, and KM3NeT \citep[18\,TeV neutrino event found in their archival data][]{Filippini22}.

In our time frame of interest (Modified Julian Date; MJD: 59540 -- 59754, decimal year: 2021.89 -- 2022.48) \ze\ was observed six times with the VLBA in radio wavelengths and with the Perkins Telescope Observatory in optical wavelengths as part of the VLBA-BU-BLAZAR/BEAM-ME programme; the data are publicly available.
\cite{Jorstad17} provide details of the observations, data calibration in both total intensity and linear polarisation, and imaging.
Each individual observation was then described by means of geometrical model-fitting within the forward modelling \textit{eht-imaging} software suite \citep{Chael16, Paraschos24c} in combination with the heuristic optimisation tools implemented in \texttt{SciPy} \citep{2020SciPy-NMeth}.
In short, geometrical model-fitting enables us to represent the complex structure of a jet in a more simplified way (in comparison with, for example, \texttt{CLEAN} component representation), by describing its morphology with a number of parametric components, for example circular Gaussian functions.
As shown in \citep{Paraschos24a}, we decreased the number of free parameters in this way (by using circular instead of elliptical Gaussian components), thus, simplifying the modelling procedure for sparse data.
All relevant parameters (such as total and linear intensity, position, EVPAs) of each Gaussian component can then be easily extracted, facilitating investigations into how variability in the jet structure and in light curves might be connected \citep[e.g.][]{Paraschos22, Paraschos23}.

Finally, we obtained the \textit{Fermi} Large Area Telescope (LAT) Collaboration \citep[see][]{Abdollahi23} weekly binned $\gamma$-ray light curve from the online repository\footnote{\url{https://fermi.gsfc.nasa.gov/ssc/data/access/lat/msl_lc/}}, which contains publicly available data.

\begin{figure}
\centering
\includegraphics[scale=0.27]{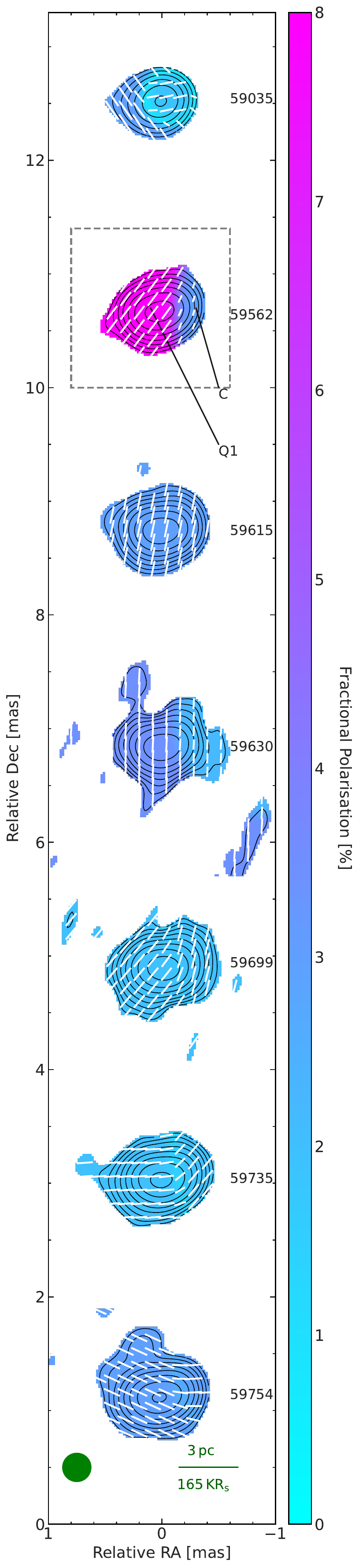}
  \caption{
    Fractional polarisation geometrical model-fit (colour) and Stokes I image (contours) of \ze\ showcasing the epochs close in time to the MJD 59500--59650 (year: 2021.78--2022.19) multi-band flare.
    The colour scale is linear, and the units are in percent.
    The dark green ellipse in the lower-left corner shows the common convolving circular beam size of \(\beam\).
    The green bar (bottom right) corresponds to a projected distance of $165 \textrm{K}R_\textrm{S}$ (Schwarzschild radii).
    Superimposed are the radio EVPAs (white sticks), showcasing the direction of the electric field, with the magnetic field direction being perpendicular to them.
    Their length is proportional to the fractional polarisation values.
    The lowest Stokes I contour cut-oﬀ at $4\sigma_\textrm{I}$ was implemented to only include high-$S/N$ areas (with \(\sigma_\textrm{I} = \rms\)).
    The contour levels are at 0.25, 0.5, 1, 2, 4, 8, 16, 32, and 64\% of the total intensity peak per epoch, the MJDs of which are denoted next to each observation.
    The core (C) and jet (Q1) components identified throughout all epochs are denoted as an example in the second epoch (indicated by dash-lined grey box), which is closest in time to the neutrino detection.
    The radio EVPAs in C start out along the bulk jet flow (east-west direction).
    As the shock front progresses downstream, they exhibit a clear turn, becoming perpendicular to the bulk jet flow, before returning to their initial quiescent (east-west) state in the last image.
    }
   \label{fig:VLBI}
\end{figure}

\section{Results}\label{sec:Results}

In order to best fit the morphology of \ze\ we applied models containing one, two, and three components to the data. 
The fit of a one component model was inadequate, yielding $\upchi^2_\textrm{tot}\gg1$.
The three component case also corresponds to the result published by \cite{Kim25} within the \texttt{DIFMAP/CLEAN} algorithm framework, where the authors reported the ejection of another jet component, thus, fitting the entirety of the core and jet with three components.
Our analysis, utilising the aforementioned less-supervised, regularised maximum likelihood (RML) approach, showed that the structure of \ze\ is best approximated by two components: a core component labelled `C' (assumed to be stationary and set at zero) and a jet component labelled `Q1'.
A three component model resulted in over-fitting in this framework (we implemented a cut-off of $\upchi^2_\textrm{tot}<0.5$).
The $\upchi^2_\textrm{tot}$ for each model-fitted epoch is shown in the last column of Table~\ref{table:Params}.

Our decision to identify the core as the western component was motivated by the long term monitoring of the source, which has ejected components moving eastwards for decades \citep[see e.g.][]{Britzen10}.
This double structure persists throughout the time frame in question and remains quasi-stationary.
The morphological evolution of \ze\ is presented in Fig.~\ref{fig:VLBI}.
Specifically, we show the \texttt{CLEAN} \citep{Shepherd95} images in contours, and the fractional linear polarisation and EVPAs of C and Q1 in colour and with white sticks, respectively.
The parameters of these two components are listed in Table~\ref{table:Params}, which our further analysis is based on.
We also include the last available epoch before the neutrino event and multi-band flare, from MJD 59035 (year 2020.35).
Even though its temporal separation from the time frame of interest is significant, this epoch serves the very useful purpose of revealing the jet parameters before the multi-band flare onset.
Since we cannot definitively assert that the jet components needed to fit the jet in the first epoch correspond to the same ones we detected during the time frame of interest, we labelled them `Q01' and `Q02' respectively.

\begin{figure*}
\centering
\includegraphics[scale=0.75, trim={3.5cm, 0cm, 3.5cm, 0cm}]{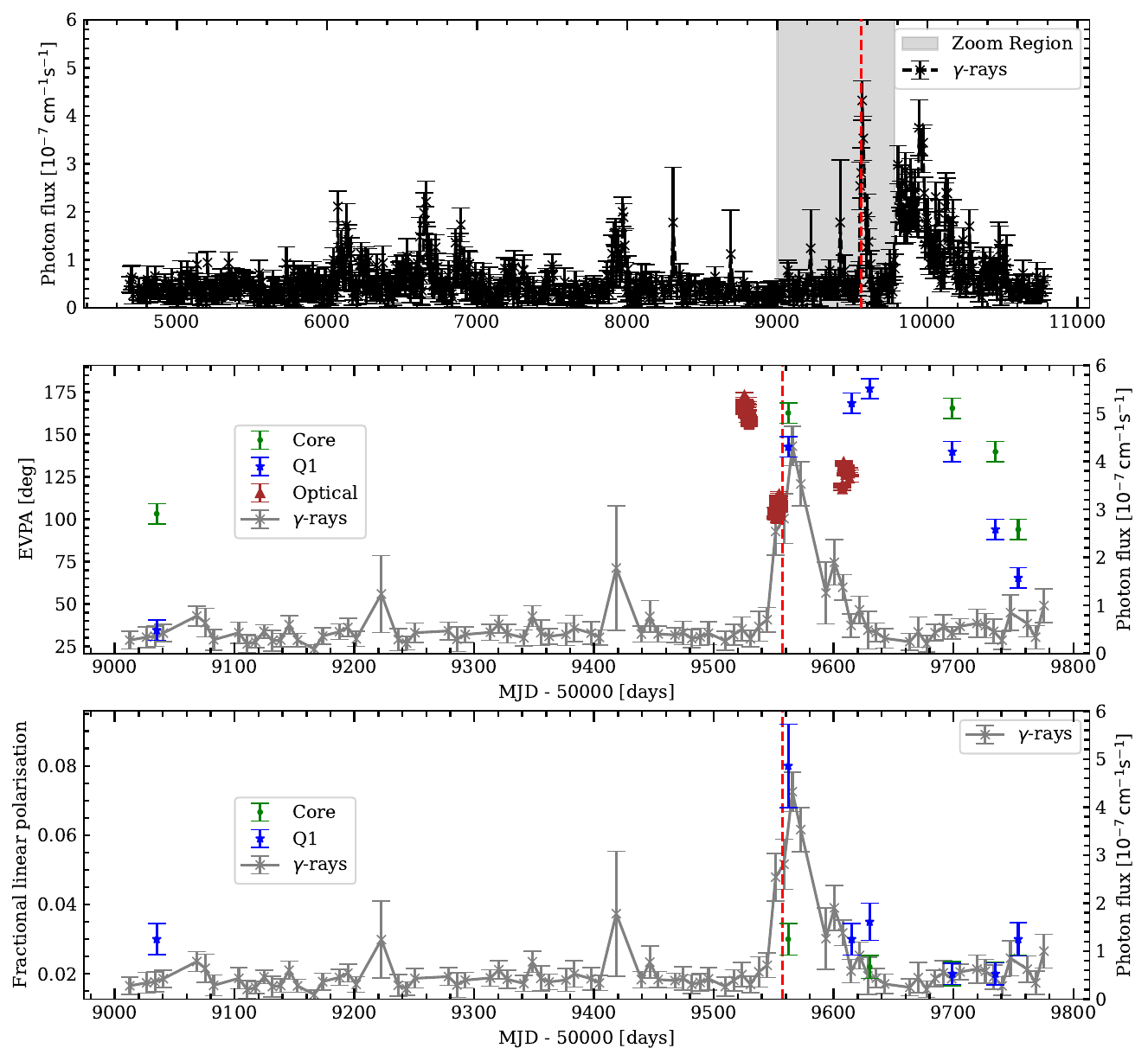}
  \caption{
    \ze\ component EVPAs (radio) and flux density, optical EVPAs, and the \textit{Fermi}-LAT $\gamma$-ray light curve.
    In the top panel, the grey-shaded area marks the period in question, when the prominent $\gamma$-ray flare occurred.
    Since then, the source has also exhibited a prolonged flaring activity, albeit without further neutrino events reported so far.
    The middle panel shows a zoomed-in view of the $\gamma$-ray flare time frame (grey-shaded area of the top panel), close-in-time to the neutrino detection.
    Additionally, the radio EVPAs in C (green) and Q1 (blue) show a clear rotation, with the core radio EVPAs being aligned to the bulk jet flow before and after the $\gamma$-ray flare.
    Analogous behaviour is observed for optical EVPAs (shown with brown markers).
    The vertical, dashed, red line shows the time of arrival of the neutrino. 
    The bottom panel is similar to the middle one except that here the green and blue markers reflect the fractional linear polarisation of C and Q1 respectively.
    }
     \label{fig:gamma}
\end{figure*}

\begin{table*}
\caption{Summary of the component parameters.}
\label{table:Params}
\centering
\begin{tabular}{cccccccc}
ID & Obs. [MJD - 50000 / Dec. year] & F0 [Jy] & FWHM [mas] & Position [mas, mas] & Frac. pol [\%] & EVPA [deg] & $\upchi^2_\textrm{tot}$\\
\hline\hline
 C   &  9035 / 2021.51 &  $0.18\pm0.03$ &  $0.08\pm0.05$ &  [0.00, 0.00]  &  $2.0\pm0.3$ &  $103\pm6$ & 0.69 \\
 C   &  9562 / 2021.95 &  $0.47\pm0.07$ &  $0.12\pm0.05$ &  [0.00, 0.00]  &  $3.0\pm0.5$ &  $163\pm6$ & 1.00 \\
 C   &  9615 / 2022.10 &  $0.50\pm0.07$ &  $0.12\pm0.05$ &  [0.00, 0.00]  &  $3.0\pm0.5$ &  $169\pm6$ & 0.69 \\
 C   &  9630 / 2022.14 &  $0.47\pm0.07$ &  $0.10\pm0.05$ &  [0.00, 0.00]  &  $2.2\pm0.3$ &  $177\pm6$ & 0.90 \\
 C   &  9699 / 2022.33 &  $0.93\pm0.14$ &  $0.12\pm0.05$ &  [0.00, 0.00]  &  $2.0\pm0.4$ &  $166\pm6$ & 0.82 \\
 C   &  9735 / 2022.42 &  $0.82\pm0.12$ &  $0.12\pm0.05$ &  [0.00, 0.00]  &  $2.0\pm0.3$ &  $140\pm6$ & 1.31 \\
 C   &  9754 / 2022.48 &  $0.55\pm0.08$ &  $0.12\pm0.05$ &  [0.00, 0.00]  &  $3.0\pm0.5$ &  $94\pm6$ & 1.64 \\
 Q01 &  9035 / 2021.51 &  $0.12\pm0.02$ &  $0.16\pm0.05$ &  [0.11, -0.06] &  $3.0\pm0.5$ &  $35\pm6$ & 0.69 \\
 Q1  &  9562 / 2021.95 &  $0.37\pm0.06$ &  $0.13\pm0.05$ &  [0.17, 0.00]  &  $8.0\pm1.2$ &  $143\pm6$ & 1.00 \\
 Q1  &  9615 / 2022.10 &  $0.43\pm0.06$ &  $0.12\pm0.05$ &  [0.17, -0.04] &  $3.0\pm0.5$ &  $169\pm6$ & 0.69 \\
 Q1  &  9630 / 2022.14 &  $0.52\pm0.08$ &  $0.13\pm0.05$ &  [0.17, -0.03] &  $3.5\pm0.5$ &  $177\pm6$ & 0.90 \\
 Q1  &  9699 / 2022.33 &  $0.57\pm0.09$ &  $0.13\pm0.05$ &  [0.17, -0.04] &  $2.0\pm0.3$ &  $140\pm6$ & 0.82 \\
 Q1  &  9735 / 2022.42 &  $0.74\pm0.11$ &  $0.13\pm0.05$ &  [0.17, -0.03] &  $2.0\pm0.3$ &  $94\pm6$ & 1.31 \\
 Q1  &  9754 / 2022.48 &  $0.74\pm0.11$ &  $0.15\pm0.05$ &  [0.17, -0.06] &  $3.0\pm0.5$ &  $65\pm6$ & 1.64 \\
 Q02 &  9035 / 2021.51 &  $0.01\pm0.01$ &  $0.95\pm0.05$ &  [0.22, 0.21]  &  $3.0\pm0.5$ &  $34\pm6$ & 0.69 \\
\hline
\end{tabular}
\begin{tablenotes}
    \item Starting from left to right, we present the component identification (ID), day of observation (Obs.), flux density (F0), full width at half maximum (FWHM), position, fractional linear polarisation ($m$), EVPA of each component, and the total $\upchi^2$ for each epoch.
\end{tablenotes}
\end{table*}

We find that the jet and core were in a quiescent state before the multi-band flare and neutrino event.
The core radio EVPAs were aligned to the bulk jet flow and the overall fractional linear polarisation rather low, of the order of $m\sim2.5\%$.
Then, immediately after the neutrino detection and while the $\gamma$-ray flare was developing, the fractional linear polarisation increased in the jet to $m_{\textrm{Q}_1}\sim8\%$, and the radio EVPAs in both the jet and core are almost perpendicular to the bulk jet flow.
As time progresses the fractional linear polarisation decreases again in both the jet and core, reaching its pre-flare values and the radio EVPAs turn until they are finally aligned again to the bulk jet flow in the core (i.e. its quiescent state).
Simultaneously, the flare was detected across the electromagnetic spectrum, as reported in multiple works \citep[e.g.][]{Acharyya23, Sahakayan23}. 
Interestingly, during the $\gamma$-ray flare coincident with the neutrino arrival we observe an optical polarisation angle rotation of $\Delta\theta_\textrm{opt}\sim80^\circ$ (Fig.~\ref{fig:gamma}). 
Unfortunately the optical observations stopped before reaching the $\gamma$-ray peak, which happens to be the highest peak observed in \ze\ to-date. 
Therefore, the rotation likely lasted longer and had a larger amplitude. 
Given the well-established statistical connection between polarisation angle rotations and $\gamma$-ray flares \citep{Blinov15, Blinov18}, and the clear delayed rotation of the polarisation vectors at 43\,GHz, we conclude that the optical rotation at the time of the neutrino arrival was a physical event connected to the $\gamma$-ray flare and not a result of the magnetic field's random variations.

\section{Discussion}\label{sec:Discussion}

\subsection{Radio EVPA rotation}\label{ssec:Rotation}

Our observed rotation of the radio EVPAs in the core and jet is a phenomenon that has been described before in other blazars.
\cite{Liodakis22} showed that a shock front propagating through a stationary recollimation shock (shock-shock interaction) provides the necessary environment to produce this radio EVPA geometry and results in the increase in fractional linear polarisation \citep[see also][and bottom panel of Fig.~\ref{fig:gamma}]{Kramer24}.
Then, in the framework described in the works of \cite{Daly88}, \cite{Marscher85}, and \cite{Marscher02}, a travelling shock front (shock-in-jet model) enhances particle acceleration by amplifying shear processes or interactions with the surrounding medium of ambient magnetic field lines, initially aligned with the jet axis.
An accompanying flare is also expected, as was in fact observed for \ze.
Interestingly, a similar morphology has also been observed in other sources, as presented in \cite{Larionov13}, \cite{Liodakis20}, \cite{Kouch25a}, and \cite{Paraschos25a}.
Specifically in the latter case, the author describes a shock front moving downstream, without a new component appearing in the jet, as it is also the case for \ze.
We note, however, that in the analysis of the same source presented in \cite{Kim25}, the authors find that a superluminal component is ejected during the neutrino emission time frame, further strengthening our case.

\subsection{Jet component -- neutrino emission coincidence}

Brightness enhancements, commonly referred to as jet components, are a frequent feature of astrophysical jets. 
They can arise from shocks, plasma blobs propagating within the bulk jet flow, or geometric effects such as twisted filamentary structures bent toward our line of sight \citep[e.g.][]{Traianou24, Paraschos25b}.
A possible connection between such jet components and neutrino events has already been presented for TXS\,0506+056 \citep[for example in][]{Kun19, Ros20, Sumida22}, as well as other sources \citep[e.g.][]{Nanci22, Nanci23, Eppel23}.
The flaring activity seen in radio total variability light curves in these works, which is thought to be potentially connected with the neutrino events, could stem, for example, from the emergence of new components from the core,  helical trajectories, spine-sheath structures \citep{Ghisellini05, Paraschos24b}.

In this framework we investigated the jet morphology of \ze\ in both total intensity and fractional linear polarisation coincident with the IceCube-211208A neutrino event.
Fortunately, the jet of \ze\ has been studied for decades, providing a wealth of insights into its structural changes using polarised light \citep{Gabuzda94, Gomez99, Agudo06} and kinematics \citep{Marscher80, Gomez01, Britzen10}.
Overall, the jet seems to be oscillating between active and quiescent phases.
In the active phases the jet exhibits superluminal motion and a curved trajectory.
In the quiescent ones its morphology remains rather constant, with the jet components needed to describe it remaining quasi-stationary.
It is in this latter stage that the neutrino event was detected in \ze, which is expected based on the discussion presented in \cite{Plavin21}.
There the authors postulate that for a neutrino in the TeV to PeV range to be emitted via a proton--photon interaction, mildly relativistic shocks near the jet base need to provide the necessary energy to accelerate protons.
This is exactly what we are observing in \ze, as revealed by the radio and optical EVPA rotation and linear polarisation increase during the multi-band flare that coincided with the neutrino event.

The second ingredient required for the proton--photon interaction are the target photons to be hit by the accelerated protons.
As shown in the spectral energy distribution studies related to the multi-band flare of \ze, an external photon field can better model the multi-messenger spectrum of the source \citep[e.g.][]{Acharyya23, Sahakayan23, Prince24, Omeliukh25}.
Considering then again the jet's active phase, which is characterised by numerous twists and turns, a dense surrounding medium could be responsible for this behaviour.
This ambient medium could then potentially be the target photon source \citep[see also][for a similar discussion about TXS\,0506+056]{Reimer19}.
\cite{Kim25} argue, however, that models less sensitive to external radiation might be more preferred, assuming that the neutrino emission region is separated from the core by almost one parsec.

Alternatively, as shown in \cite{Mastichiadis21}, protons interacting with their own synchrotron radiation can create neutrinos via photomeson interactions.
The telltale sign of this scenario is an X-ray flare, which is observed in \ze.
In that case, the target photon could be also internal to the jet.

As postulated in \cite{Britzen10}, an impulse related to a flare might inject enough kinetic energy into the \ze\ jet to straighten it out, thus, explaining the repeated transition between the two phases every $\sim9$\,years.
Such flaring activity in $\gamma$-rays has been shown to potentially be connected to neutrino events, as well as rotations of the optical polarisation plane \citep[e.g.][]{Blinov21, Novikova23, Blinov25}, which is exactly what we observe in the middle panel of Fig.~\ref{fig:gamma}.
The combination of this optical and radio (delayed by a few months) EVPA rotation strengthens our case for polarised flux being a probe of neutrino emission in blazars.

Based on the results presented in that same study by \cite{Britzen10}, \ze\ entered quite recently, and is currently undergoing, its quiescent phase, also corroborated by our results presented in Fig.~\ref{fig:VLBI}.
Specifically, using the reported $\sim9$\,year timescale and the fact that in late 2008 the jet was in its previous quiescent state, we estimate that in the 17 years since then, \ze\ went through another active phase, after which it started its current quiescent state, around the year 2020.
Thus, under the assumption that neutrino events are favoured by a quasi-stationary jet morphology, we predict that the next neutrino association with \ze\ will occur in $\sim18$\,years time, during its next quiescent period.
This prediction does of course not preclude further neutrino detections during its current quiescent phase, which should end and turn into active by the end of the decade.

\section{Conclusions} \label{sec:Conclusions}

In this study we have investigated the association between the jet morphology changes of \ze\ and the accompanying flare detected in multiple bands, and have discussed how they might be related to a contemporary neutrino event.

Our work can be summarised as follows:
We find that during the multi-band flare coincident with neutrino event IceCube-211208A, the jet of \ze\ also showed high levels of fractional linear polarisation. 
The radio EVPAs in the core rotated, initially being perpendicular to the bulk jet flow, then turning parallel to it, before ending up in the initial, perpendicular orientation.
This morphological change is consistent with a shock propagating through the jet.
Similarly, the optical EVPAs exhibit a rotation of $\Delta\theta = 80^\circ$. 
Since the optical observations stopped abruptly, we conclude that the duration and amplitude of the rotation was in fact larger.
Such a shock is a proposed mechanism for proton acceleration capable of producing neutrinos via the proton-photon interaction process. 
Close-in-time to the neutrino event, the \ze\ jet is best described quasi-stationary components, consistent with what is expected from the literature for blazars associated with neutrinos \citep{Plavin21}.
Whilst \ze\ was in this quiescent state in 2021, it also exhibits periods of high activity, in which the jet is highly curved.
A dense ambient medium surrounding the jet and causing these bends would be a natural explanation for the source of the target photons.
Another target source could be photomeson interactions, as suggested by the temporary coincident X-ray flare.

While our work presented here provides possible insights into the association between a neutrino and a blazar, we have been operating under the assumption that such a connection exists in the first place.
However, we are still in the infancy of multi-messenger astronomy with neutrinos, awaiting more concrete proof that this connection does in fact exist \citep[see also][for a description of the necessary conditions]{Liodakis22b}.
Synergy between future instruments, like the next-generation Event Horizon Telescope \citep[ngEHT; see][]{Kovalev23}, Square Kilometer Array \citep[SKA; see][]{Paragi15}, and next-generation Very Large Array \citep[ngVLA; see][]{Murphy18} will be crucial to answer long-standing open questions, such as how and where neutrinos are produced, and what the particle acceleration mechanisms are, by resolving the different possible jet acceleration regions and monitoring the structural changes of the jets.

\begin{acknowledgments}
      We thank the anonymous referee for their valuable comments which greatly improved this manuscript for the insightful discussions.
      We are sincerely grateful to Svetlana Jorstad for the provision of the optical polarisation observations and Sergio Dzib for fruitful discussions.
      This research is supported by the European Research Council advanced grant “M2FINDERS - Mapping Magnetic Fields with INterferometry Down to Event hoRizon Scales” (Grant No. 101018682). 
      I. L. was funded by the European Union ERC-2022-STG - BOOTES - 101076343. Views and opinions expressed are however those of the author(s) only and do not necessarily reflect those of the European Union or the European Research Council Executive Agency. Neither the European Union nor the granting authority can be held responsible for them.
      This study makes use of VLBA data from the VLBA-BU Blazar Monitoring Program (BEAM-ME and VLBA-BU-BLAZAR;
      \url{http://www.bu.edu/blazars/BEAM-ME.html}), funded by NASA through the Fermi Guest Investigator Program. The VLBA is an instrument of the National Radio Astronomy Observatory. The National Radio Astronomy Observatory is a facility of the National Science Foundation operated by Associated Universities, Inc.
      This study was based in part on observations conducted using the 1.8\,m Perkins Telescope Observatory (PTO) in Arizona, which is owned and operated by Boston University.
      This research has made use of the NASA/IPAC Extragalactic Database (NED), which is operated by the Jet Propulsion Laboratory, California Institute of Technology, under contract with the National Aeronautics and Space Administration. 
      This research has also made use of NASA's Astrophysics Data System Bibliographic Services. 
\end{acknowledgments}

\begin{contribution}

G.~F.~P. led the writing of the paper and the VLBI analysis.
E.~T., L.~C.~D., I.~L., and E.~R. contributed to the text and the interpretation of the results.
\end{contribution}

\facilities{Very Long Baseline Array, Perkins Telescope Observatory}

\software{numpy \citep{Harris20}, scipy \citep{2020SciPy-NMeth}, matplotlib \citep{Hunter07}, astropy \citep{2013A&A...558A..33A, 2018AJ....156..123A}, and 
{Uncertainties: a Python package for calculations with uncertainties.}
          }

\bibliography{sample701}{}
\bibliographystyle{aasjournalv7}

\end{document}